\documentclass[aps,pre,twocolumn]{revtex4}
\usepackage{amsmath,amssymb}
\usepackage{color,graphicx}

\begin{document}

\title{ L\'evy flights in the infinite potential well as the hypersingular Fredholm problem}
\author{Elena V. Kirichenko, Piotr Garbaczewski, Vladimir  Stephanovich and  Mariusz  \.{Z}aba}
\affiliation{Faculty of Mathematics, Physics and Informatics,
University of Opole, 45-052 Opole, Poland}
\date{\today }
\begin{abstract}
We study L\'evy flights {{with arbitrary index $0< \mu \leq 2$}} inside a potential well of infinite depth. Such problem appears in many physical systems ranging from stochastic interfaces to fracture dynamics and multifractality in disordered quantum systems. The major technical tool is a transformation of  the  eigenvalue problem for initial fractional Schr\"odinger equation into that for Fredholm integral equation with hypersingular kernel. The latter equation is then solved by means of expansion over the complete set of orthogonal functions in the domain $D$, reducing the problem to the spectrum of a matrix of infinite dimensions. The eigenvalues and eigenfunctions are then obtained numerically with some analytical results regarding the structure of the spectrum.
\end{abstract}
 \maketitle

\section{Introduction}
To describe the complex behavior of disordered systems without using Gaussian approximation, the stochastic processes, called L\'evy flights, are commonly utilized \cite{zas,hu,garb}. The stochastic trajectories of L\'evy flights alternate between some continuous motions and jumps (sometimes extremely long) and hence do not obey Gaussian statistics \cite{gk,dsu,mk}. The length of these jumps obeys to so-called L\'evy stable distributions  with a power-law tails, which decay much slower then Gaussian ones. This yields the divergence of already second moment of such distributions. Contrary to ordinary diffusion, described by Gaussian distributions, the above jump-type discontinuous motions are commonly attributed as anomalous diffusion \cite{mk,dsu}. It turns out that L\'evy stable distributions and L\'evy flights are relevant to many physical \cite{zas1,ux1,ux2,lc,mc}, chemical, biological \cite{bi1,bi2,bi3} and socio-economic \cite{ek1,ek2,ek3} systems. Prominent physical examples are subrecoil laser cooling of trapped atoms \cite{lc}, energy exchange in Landau-Teller model of molecular collisions \cite{mc} and so-called multifractality of the wave functions in the disordered quantum systems \cite{dq1,dq2}.

It is well-known, that the concentration $n(x,t)$ of particles performing L\'evy flights satisfies in its simplest form a diffusion equation where the Laplacian operator is replaced by a fractional derivative
\begin{equation}\label{cw1}
\frac{\partial n(x,t)}{\partial t}=- |\Delta |^{\mu /2}n(x,t),
\end{equation}
where $|\Delta |^{\mu /2}$ (L\'evy index $0<\mu \leq 2$) is a fractional Laplacian of order $\mu /2$, restricted to 1D case \cite{sz97} so that at $\mu=2$ we recover the ordinary Laplace operator \cite{fk1,fk2}. {{The explicit form of this operator reads
 \begin{eqnarray}
 &&|\Delta |^{\mu /2} f(x)\, =-A_\mu \int_R {\frac{f(u)- f(x)}{|u-x|^{1+\mu }}}\, du, \label{cw2} \\
 &&A_\mu={\frac{\Gamma (\mu +1) \sin(\pi \mu/2)}{\pi }},
\end{eqnarray}
which shows}} that this operator is spatially nonlocal, becoming an issue if confronted with  a priori imposed boundary conditions. This is irrespective of whether we are interested in L\'{e}vy processes with absorption (killing) at the boundaries, or in so-called fractional quantum mechanics.

Namely, at the unbounded domains, the fractional Laplacian \eqref{cw2} is most easily defined by its Fourier transform
\begin{equation} \label{cw3}
\frac{1}{2\pi} \int_{-\infty}^\infty   |k|^\mu f(k)
  e^{-\imath kx}dk  \equiv - \frac{\partial _{\mu }f(x)}{\partial |x|^{\mu }}  =  |\Delta |^{\mu /2}f(x),
\end{equation}
while on bounded domains $D\subset R$ ($R$ is a real axis) the Fourier transform \eqref{cw3} is no longer  operational  \cite{GS,ZG,pgar,zg,olk,ZRK,getoor}, see specifically \cite{pgar,zg}. Transformation \eqref{cw3} permits to solve many problems related to fractional diffusion and fractional quantum mechanics in $k$ space on unbounded domains \cite{dsu, gs09, gs11}, while on the bounded ones this method fails, making problem nontrivial.

In this paper, we investigate the L\'evy flights {{of arbitrary index $0< \mu \leq 2$}} confinement by the infinite potential well, which arises naturally in the context of so-called first-passage problems \cite{fp1,fp2}. We show that this problem is equivalent to that of fractional quantum mechanics of a particle in a potential well of infinite depth \cite{luchko,iomin}. We solve this problem by further reducing the corresponding fractional Schr\"odinger equation to the Fredholm integral equation with hypersingular kernel. The latter equation has been solved with arbitrary accuracy (for eigenstates and eigenfunctions) by the expansion over the (infinite) complete set of orthogonal functions, which in the case of above potential well are trigonometric functions. {{This expansion is suitable for any L\'evy index $0< \mu \leq 2$, although many results are obtained for
so-called ultrarelativistic or Cauchy case $\mu=1$. This case corresponds to zero mass ($m=0$) case of the relativistic Hamiltonian ${\cal H}=\sqrt{-\hbar^2c^2\Delta+m^2c^4}-mc^2$ ($c$ is the velocity of light, $\Delta$ is ordinary Laplacian) and thus is physically sound \cite{ZG}-\cite{ZRK}). Note also purely mathematical literature \cite{getoor,riesz,hadamard,K,KKMS,dyda} in the context of the case $\mu=1$.}}
We show, that our algorithm, consisting in the expansion of the solution over the suitable set of orthogonal functions, permits to attack successfully virtually any problem of so-called fractional quantum mechanics.

\section{Fredholm integral equation for the spectrum}
{{We consider the fractional Schr\"odinger equation
\begin{equation}\label{fr1}
\left[-|\Delta |^{\mu/2} + V(x)\right]\psi(x)=E \psi (x)
\end{equation}
where
\begin{equation} \label{fr1a}
V(x)=\left\{\begin{array}{c}0, \ x\in [-1,1] \\ \infty, \quad {\rm {otherwise}},
\end{array}\right.
\end{equation}
which implies that $\psi(x) =0$ for $|x| \geq 1$ and defines the infinitely high "walls" of the potential well in the points $x=\pm 1$. Now, the whole real axis $R$ can be divided into the regions inside $-1 \leq x \leq 1$ and outside well \eqref{fr1a} to get:
\begin{eqnarray}
&&|\Delta |_D^{\mu/2}\psi(x)=-A_\mu \int_{-\infty}^\infty \frac{\psi(u)-\psi(x)}{|u-x|^{1+\mu}}du=\nonumber \\
&&=-A_\mu\left[\int_{-\infty}^{-1}+\int_{-1}^1+\int_1^\infty\right]\frac{\psi(u)-\psi(x)}{|u-x|^{1+\mu}}du.\label{fr1b}
\end{eqnarray}
The symbolic integration signs in the square brackets in \eqref{fr1b} mean the sum of corresponding
integrals. Now we make note of the fact that for regions outside the well $\psi(u)=0$ so that we have
\begin{eqnarray}
 &&|\Delta |_D^{\mu/2}\psi(x)=-A_\mu \int_{-1}^1\frac{\psi(u)du}{|u-x|^{1+\mu}}+\Delta I_\mu, \label{fr1c} \\
 &&\Delta I_\mu=A_\mu \psi(x)\left[\int_{-\infty}^{-1}+\int_{-1}^1+\int_1^\infty\right]\frac{du}{|u-x|^{1+\mu}}.\nonumber
\end{eqnarray}
It can be shown that with respect to definition of modulus ($|u-x|=u-x$ if $u>x$ and $x-u$ if $u<x$) and above division of real axis $R$ into three subintervals, the integral $\Delta I_\mu$ is defined by the values of antiderivative at infinities. These values are zero except the case $\mu=0$, where they are logarithmically divergent. This yields $\Delta I_\mu \equiv 0$ for $0< \mu \leq 2$ so that the desired integral equation acquires the form
\begin{equation} \label{ii2a}
- A_\mu \int_{-1}^{1}\frac{\psi(u)dt}{|u-x|^{1+\mu}}=E \psi (x).
\end{equation}
The equation \eqref{ii2a} is the Fredholm integral equation, which we are going to solve below. We will show that at $\mu=2$ our solution recovers the case of the infinite potential well with ordinary Laplacian \cite{landau3}.
Note that for $\mu=1$ the integral in \eqref{ii2a} refers to the so-called Hadamard finite part of singular integrals, extensively employed in the works of crack propagation in solids \cite{hadamard,K,KKMS,dyda,stein,kaya, ang,chan}.

Note also that the spectral problem \eqref{ii2a} is the  homogeneous Fredholm equation  with a hypersingular symmetric kernel  $K(t,x) = A_\mu |u-x|^{-1-\mu}$. If the kernel of Eq. \eqref{ii2a} is nonsingular (i.e. such that $\int_a^b\int_a^b K^2(x,t) dx dt < \infty $), then this equation obeys so-called Fredholm alternative \cite{pol}: either $E$ (or $\lambda=1/(\pi E)$) is its eigenvalue and $\psi$ is eigenfunction or the equation has a trivial solution $\psi(x)=0$. Also, for nonsingular kernel, the number of eigenstates is discreet and finite \cite{pol} and exactly for the equation \eqref{ii2a} with nonsingular kernel its eigenfunctions are $\sin (n\pi x)$ and $\cos (n\pi x/2)$, i.e. they correspond to the case of infinite well in ordinary quantum mechanics \cite{landau3}. On the other hand, for the case of singular kernels, the solution of the spectral problem (if in existence) has an infinite (although discreet) number of eigenstates \cite{pol}.

One more remark is in place here. As we will see below, the best way to solve the integral equation \eqref{ii2a} is to expand its solution over the complete set of orthogonal functions. In our view, the best choice of such set is the eigenfunctions of the corresponding ordinary (i.e. that with ordinary Laplacian) quantum mechanics. In other words, we can claim, that the fractional derivative in corresponding quantum mechanical problem "mixes" all the eigenstates of that with ordinary Laplacian. This means, for instance, that even ground state wave function for $\mu \neq 2$ is indeed an infinite superposition of the functions, corresponding to $\mu=2$. Below we are going to realize this algorithm.}}

\section{The solution of the integral equation}

Now we are going to solve the integral equation \eqref{ii2a}, i.e. to deduce the  eigenfunctions and eigenvalues  of the nonlocal {{operator $|\Delta |^{\mu/2}_D$. As we have mentioned above \cite{pol},
there are no systematic methods (even numerical) of solution of integral equations with singular (or hypersingular) kernels. Along the lines of above scenario, below we suggest an effective algorithm of such solution, based on the "mixture" of the quantum states of the infinite potential well with ordinary Laplacian, i.e. that for $\mu=2$. More presicely, we are looking for the solution as an expansion over the appropriate complete set of orthogonal functions, which in our case turn out to be trigonometric Fourier series.}} Our algorithm permits to obtain the eigenfunctions and eigenvalues of the problem \eqref{ii2a} with arbitrary accuracy by reducing it to the eigenproblem of the infinite matrix. Our method permits also to obtain approximate analytical expressions for eigenvalues and several first eigenfunctions.

Our algorithm is based on the following assumptions:

1. Based on standard quantum mechanical infinite well experience \cite{landau3} and  previous attempts to solve the L\'{e}vy - stable infinite well  problem \cite{K,KKMS} and \cite{ZG,pgar,zg} we can safely classify  eigenfunctions to be odd or  even. The oscillation theorem \cite{landau3} appears to be valid here so that the ground state wave function has no nodes (intersections with $x$ axis), first excited state has one node, second one has two nodes etc. So, our even states can be labeled by quantum numbers $k=$ 0,2,4,6,... while odd states by  $k=$1,3,5,....

2. Similar to ordinary quantum mechanics \cite{landau3}, the Hilbert space of the system can be interpreted as a direct sum of odd and even subspaces, equipped with corresponding orthonormal sets of functions in the interval [-1,1].

3. {{As the complete set of eigenfunctions of the ordinary ($\mu=2$) infinite well \cite{landau3}  consists of standard trigonometric functions, we will look for the eigenfunctions of the problem \eqref{ii2a} in the form of trigonometric series.}}

4. The  even basis system in $L^2(D)$ is composed of cosines
\begin{eqnarray}\label{so1}
\varphi_k(x)=\cos \frac{(2k+1)\pi x}{2},\nonumber \\
\int_{-1}^1\varphi_k(x)\varphi_l(x)dx=\delta_{kl},
\quad  k\geq 0,
\end{eqnarray}
where $\delta_{kl}$ is the  Kronecker delta.
 For the odd  basis system  we  take  the sines
\begin{equation}\label{so2}
\chi_k(x)=\sin k\pi x,\quad   \int_{-1}^1\chi_k(x)\chi_l(x)dx=\delta_{kl},\quad  k\geq 1.
\end{equation}

5.  We look for eigenfunctions of  $|\Delta |^{1/2}_D$  separately in  odd and even Hilbert subspaces  of $L^2(D)$.

 Presuming that the Fourier (trigonometric)  series converge, for even functions we have
\begin{equation}\label{so3}
\psi_e(x)=\sum_{k=0}^\infty a_{k \mu}\cos \frac{(2k+1)\pi x}{2},
\end{equation}
while for odd functions
\begin{equation}\label{so3a}
\psi_o(x)=\sum_{k=1}^\infty b_{k \mu}\sin k\pi x.
\end{equation}
To avoid confusion, we point out that the standard numbering of   overall  infinite well  eigenfunctions   begins with $n=1$ rather then from  $k=0$  (even case) or $k=1$ (odd case) as we  have assumed above. We need to have  a clear discrimination between    sine (odd) and cosine (even) Fourier series expansions.  The final outcomes will be re-labeled in terms of consecutive integers $n=1,2,...$.

\subsection{Even subspace}

{{In this case we substitute the function $\psi_e(x)$ \eqref{so3} into  \eqref{ii2a} to obtain
\begin{equation}\label{so4}
\sum_{k=0}^\infty a_{k \mu} f_{k \mu}(x)=E \sum_{k=0}^\infty a_{k \mu} \cos \frac{(2k+1)\pi x}{2},
\end{equation}
where
\begin{equation}\label{so4a}
f_{k \mu}(x)=-A_\mu \int_{-1}^1\frac{\cos \frac{(2k+1)\pi u}{2}}{|u-x|^{1+\mu}}du.
\end{equation}
It can be shown that the integrals \eqref{so4a} are convergent for any $0<\mu\leq 2$. They can be exactly reduced to the form, which does not contain removable divergences
\begin{eqnarray}
&&f_{k \mu}(x)=-\frac{A_\mu\lambda_k^\mu}{\mu}\Biggl\{\sin \lambda_kx\int_{\lambda_{k-}}^{\lambda_{k+}}u^{-\mu}\cos u \ du-\nonumber \\
&&-\cos \lambda_kx\Biggl[\int_0^{\lambda_{k-}}u^{-\mu}\sin u \ du
+\int_0^{\lambda_{k+}}u^{-\mu}\sin u \ du\Biggr]\Biggr\}, \nonumber \\
&&\lambda_k=\frac{\pi}{2}(2k+1),\ \lambda_{k\pm}=  \lambda_k(1\pm x).\label{so4b}
\end{eqnarray}
Note that the integrals in square brackets of \eqref{so4b} are convergent at $u=0$ for all $0<\mu<2$ (we have integrable feature like $\int u^{1-\mu}du=u^{2-\mu}/(2-\mu)$ ), while the divergence at $\mu=2$ is compensated by zero of $A_{\mu \to 2} =2-\mu$.

The expression \eqref{so4b} permits to represent functions $f_{k 1}(x)$ at $\mu=1$ through sine Si$(x)$ and cosine Ci$(x)$ integral functions \cite{GR,abr}
\begin{eqnarray}
&&f_{k 1}(x)=\frac{\lambda_k}{\pi}\Biggl\{\sin \lambda_kx\Bigl[{\rm{Ci}}\ \lambda_{k-}-{\rm{Ci}}\ \lambda_{k+}\Bigr] + \nonumber \\
&&+\cos \lambda_kx\Bigl[{\rm{Si}}\ \lambda_{k-}+ {\rm{Si}}\ \lambda_{k+}\Bigr] \Biggr\},\label{so6}
\end{eqnarray}

Note that some integrals in expression \eqref{so4b} as well as the functions Ci$\lambda_{k\pm}$ are singular at $x \to \pm 1$ \cite{abr}. Nonetheless, this singularity turns out to be removable by subsequent integration with
$\varphi_k(x)$ \eqref{so1} so that the resulting matrix elements are finite, see below.

Now we multiply both sides of the  equation \eqref{so4} by $\varphi_i(x)$ \eqref{so1}  and integrate from $-1$ to $1$ with respect to the  orthonormality of $\varphi_i(x)$.  The  equation  \eqref{so4} is now replaced by an (infinite) matrix eigenvalue problem
\begin{eqnarray}
&&\sum_{i,k=0}^\infty a_{k \mu} \gamma_{\mu ki}=E a_{i \mu},\nonumber \\
&&\gamma_{\mu ki}=\int_{-1}^1f_{k \mu}(x)\varphi_i(x)dx,\  i,k,l=0,1,2,3,... ,\label{so7}
\end{eqnarray}
whose approximate solution can be done considering successive eigenvalue problems for finite $n\times n$ matrices. Note that the expressions for diagonal matrix elements $\gamma_{\mu ii}$ give already good approximation for corresponding eigenvalues, especially for large $i$.}}

The set \eqref{so7} is a linear homogeneous system, which, according to Kronecker-Capelli theorem,  has a nontrivial solution only if its determinant equals zero. This permits to determine the eigenvalues $E_{k \mu}$ and the coefficients $a_{k \mu}$ of the expansion \eqref{so1} as the eigenvectors, corresponding to each $E_{k \mu}$.  We calculate the integrals $\gamma_{\mu ki}$ numerically, but it turns out that some of them (for instance the diagonal  elements $\gamma_{1 ii}$ at $\mu=1$) can be evaluated analytically. The explicit forms of $f_{k \mu}(x)$ \eqref{so4b} and $\varphi_i(x)$ \eqref{so1} show that the matrix \eqref{so7} is symmetric, i.e. $\gamma_{\mu ki}=$ $\gamma_{\mu ik}$,  which means that eigenvalues are real.

{{We have for diagonal elements at $\mu=1$}}
 \begin{equation}\label{so11}
\gamma_{1kk}=-\frac{2}{\pi}+(2k+1){\rm{Si}}[\pi(2k+1)],
\end{equation}
while for couple of first non-diagonal elements $\gamma_{1ki}$:
\begin{eqnarray}
&&\gamma_{10}= \frac{6{\rm{Ci}}(\pi)-6{\rm{Ci}}(3\pi)+\ln 729}{8\pi}=\nonumber \\
&&=0.2773259,\nonumber \\
&&\gamma_{20}=-\frac{5}{24\pi}\left(2{\rm{Ci}}(\pi)-2{\rm{Ci}}(5\pi)+\ln 25\right)=\nonumber \\
&&=-0.2227035,
\nonumber \\
&&\gamma_{21}=\frac{5}{16\pi}\left(6{\rm{Ci}}(3\pi)-6{\rm{Ci}}(5\pi)+
\ln\frac{15625}{729}\right)=\nonumber \\
&&=0.3088509, \label{so12}
\end{eqnarray}
{{where for clarity we suppress first index $\mu=1$.}}

The explicit form of the matrix \eqref{so7} reads {{(for each $\mu$; we once more suppress this index)}}
\begin{equation}\label{so13}
\hat A_D=\left(\begin{array}{cccc}\gamma_{00} & \gamma_{10} & \cdots & \gamma_{n0} \\
\gamma_{10} &\gamma_{11} & \cdots & \gamma_{n1}  \\
 \vdots & \cdots & \cdots & \vdots \\
\gamma_{n0} & \gamma_{n1}  & \cdots & \gamma_{nn}
\end{array}\right).
\end{equation}
To find its eigenvalues and eigenvectors we use iterative procedure, considering partial matrices $2\times 2$, $3\times 3$ etc. The eigenvalues of  the  simplest partial matrix $2\times 2$ give the lowest order approximation of  ground state
 and {\em{second}} excited state $n=2$. The equation for associated
  eigenvalues reads:
\begin{equation}\label{so14}
\left|\begin{array}{cc} \gamma_{00}-E& \gamma_{10} \\
\gamma_{10} & \gamma_{11}-E
\end{array}\right|=0,
\end{equation}
The analytical expressions for $E_0$ and $E_2$ can be obtained  by means  of analytical formulas for $\gamma_{ik}$ \eqref{so11}, \eqref{so12}.
Although  computations are  cumbersome,  one arrives at  a reasonable (albeit still far form being sharp) approximation to  eigenvalues associated with  the ground state and second (or first even)
 excited state. Using numerical values of $\gamma_{1ik}$ \eqref{so7}, we calculate {{for $\mu=1$}}
\begin{eqnarray}
&&E_0=1.191256,\ E_2=4.411727.   \label{so15} \\
&&\psi (E_0)=(-0.996257,\ 0.086437), \nonumber \\
&&\psi (E_2)=(0.086437,\ 0.996257), \label{so16}
\end{eqnarray}
where $\psi (E_0)$ are eigenvectors, corresponding to eigenvalues $E_0$ and $E_2$.
In other words, the approximate (crude, low order) shapes of the eigenfunctions read
\begin{eqnarray}
&&\psi_0=-0.996257\cos \frac{\pi x}{2}+0.086437\cos \frac{3\pi x}{2},  \label{so17} \\
&&\psi_2=0.086437\cos \frac{\pi x}{2}+0.996257\cos \frac{3\pi x}{2},  \label{so18}
\end{eqnarray}
where $\psi_0(x)$ and $\psi_2(x)$ correspond to ground and second excited state. We note here that the reproduced  eigenvectors are $L^2(D)$ normalized, while an overall sign may be negative. Latter is  not important as the physically meaningful quantity is $|\psi|^2$.

{{The same procedure yields for $\mu=0.5$:
\begin{eqnarray}
&&E_0=0.995534,\ E_2=2.06879.   \label{so15a} \\
&&\psi (E_0)=(-0.991128,\ 0.132914), \nonumber \\
&&\psi (E_2)=(0.132914,\ 0.991128), \label{so16a}
\end{eqnarray}
and for $\mu=1.7$
\begin{eqnarray}
&&E_0=1.89053,\ E_2=13.4318.   \label{so15b} \\
&&\psi (E_0)=(-0.999647,\ 0.0265864), \nonumber \\
&&\psi (E_2)=(0.0265864,\ 0.999647), \label{so16b}
\end{eqnarray}
It is seen that with increase of $\mu$ the ground state energy decreases, while the difference between ground and excited states increases. Also, for decreased $\mu$ the situation is opposite. Below we investigate this question in more details.}}

By increasing the matrix order from $2$ to $3$, we improve the accuracy  with which  lowest states are reproduced and increase their number by one.  {{For $\mu=1$ we have for eigenenergies}}
\begin{equation}\label{so19}
E_0=1.1814891,\ E_2=4.3854565,\ E_4=7.569241.
\end{equation}
It is seen that while one more state appears,  numerical outcomes for  lowest states
are  corrected  by approximately 1\%. {{This statement is valid for all $0<\mu \leq 2$.}}

For  the  $6\times 6$ matrix {{and $\mu=1$}} we have
\begin{eqnarray}
&&E_0=1.1704897,\ E_2=4.35648331,\nonumber \\
&&E_4=7.52132, \ E_6=10.68291, \nonumber \\
&&E_8=13.845025,\ E_{10}=17.01393. \label{so20}
\end{eqnarray}

{{At the same time for $\mu=1.7$
\begin{eqnarray}
&&E_0=1.88345,\ E_2=13.394,\nonumber \\
&&E_4=32.4753, \ E_6=57.9598, \nonumber \\
&&E_8=89.2117,\ E_{10}=125.814. \label{so20a}
\end{eqnarray}

It is interesting to confront the above obtained (still crude) approximate eigenvalues with analytical expression,
obtained in Ref. \cite{K} (see also Ref. \cite{KKMS})
\begin{equation}\label{kw}
E_{n\mu}\approx \left[\frac{n\pi}{2}-\frac{(2-\mu)\pi}{8}\right]^\mu,\ n=1,2,3,...
\end{equation}
Table \ref{yt} shows such comparison for three representative values of $\mu$. It it seen a very good (with the accuracy less then 1\%) coincidence between numerical values (obtained from not small-sized $6\times 6$ matrix) and those from expression \eqref{kw}. This already demonstrates the accuracy of our method for arbitrary $\mu$. }}
\begin{table*}[t]
\begin{tabular}{|c|c|c|c|c|c|c|}
\hline
$i$ & 0 & 2 & 4 & 6 & 8 & 10 \\ \hline
$\mu=0.5$, num. &0.97976 & 2.04538 & 2.71443 & 3.24759 &3.70492 & 4.11305 \\ \hline
$\mu=0.5$, Ex. \eqref{kw} &0.990832 &2.0306 & 2.69535 & 3.22591 & 3.68078 & 4.0853 \\ \hline
$\mu=1.0$, num. &1.1704897 & 4.35648331 & 7.52131594 & 10.68291 & 13.845025 & 17.01393 \\ \hline
$\mu=1.0$, Ex. \eqref{kw}  &1.1781 & 4.31969 & 7.46128 & 10.6029 & 13.7445 & 16.8861 \\ \hline
$\mu=1.7$, num. &1.88345 &13.394 & 32.4753 & 57.9598 & 89.2117 &  125.814\\ \hline
$\mu=1.7$, Ex. \eqref{kw} &1.88732 &13.3603 & 32.3962 & 57.8252 & 89.0098 &  125.522\\ \hline
\end{tabular}
\caption{ {{The comparison of 6 lowest even eigenvalues $E_i$ for different $\mu$ obtained numerically from $6\times 6$ matrix and from approximate formula \eqref{kw}.}} }\label{yt}
\end{table*}

Obviously, while passing to higher order matrices the obtained solutions give better approximations to the "true" eigenvalues and eigenvectors of the infinite well problem.
The analysis of numerical values of matrix elements in  \eqref{so13} shows that for any $\mu$ these of diagonal elements are much larger than the  off-diagonal ones. This difference appears to be lowest for  $\gamma_{00}$.  For larger $k$ the diagonal elements grow (for instance at $\mu=1$ $\gamma_{22}\approx 4.388$), while off-diagonal  values  are close to 0.3. This means  that  diagonal elements give a fairly  good approximation for eigenvalues of the matrix \eqref{so13}, see the first row of Table \ref{rt}.

\begin{table*}[t]
\begin{tabular}{|c|c|c|c|c|c|c|}
\hline
$i$ & 1 & 2 & 3 & 4 & 5 & 6 \\ \hline
Diagonal elem. &1.21531728 &2.83630315 & 4.38766562 & 5.96864490 & 7.53320446 & 9.10820377 \\ \hline
$E_{i6x6}$ &1.1704897 & 2.780209 & 4.356483317 & 5.9397942 & 7.52131594 & 9.099426 \\ \hline
$E_{i12x12}$ &1.1644016 &2.7690111 & 4.3388792 &5.919976& 7.4952827 &  9.0725254\\ \hline
$E_{i10^4x10^4}$ &1.157791 &2.754795 & 4.3168638 & 5.892233 & 7.460284 & 9.032984  \\ \hline
$E_{i(K)}$\cite{K} Table 2 &1.1577 & 2.7547 & 4.3168 & 5.8921 & 7.4601 & 9.0328 \\ \hline
$E_{i(KKMS)}$\cite{KKMS} Eq. (11.1)&1.1577738 & 2.7547547& 4.3168010& 5.8921474 & 7.4601757 &  9.0328526 \\ \hline
$E_{i(ZG)}$\cite{ZG} Table VII & 1.1560 & 2.7534  & 4.3168 & 5.8945& 7.4658 &  9.0427\\ \hline
$E_{i(zg)}$ \cite{zg} Table III & 1.157776 & 2.754769 & 4.316837 & 5.892214  & 7.460282 & * \\ \hline
\end{tabular}
\caption{Comparative table of 6 lowest eigenvalues $E_i$ in the  Cauchy  infinite potential well, $\mu=1$.  Results for matrices of  different sizes in our approach  are compared with spectral data of  Refs. \cite{K}, \cite{KKMS}, \cite{ZG} and \cite{zg}. First six diagonal elements of the  matrix   (\ref{so13})  (expressions \eqref{so11} and \eqref{so24} respectively) are cited for comparison.   Note that  the numbering of states  follows  tradition ($i=1,2,3,4,5,6$) and refers to consecutive eigenvalues, with no reference to the parity of respective  eigenfunctions. }\label{rt}
\end{table*}

\subsection{Odd subspace}

{{We look for  eigenfunctions  in the form \eqref{so3a}. Repeating  the same steps as for the even subspace we generate the following set of equations
\begin{eqnarray}
&&\sum_{i,k=1}^\infty b_{k\mu} \eta_{\mu ki}=E b_{l\mu},\nonumber \\
&&\eta_{\mu ki}=\int_{-1}^1g_{k\mu}(x)\chi_i(x)dx,\ i,k,l=1,2,3,..., \label{so21}  \\
&&g_{k \mu}(x)=-A_\mu \int_{-1}^1\frac{\sin k\pi u}{|u-x|^{1+\mu}}du=  \nonumber \\
&&=\frac{A_\mu b_k^\mu}{\mu}\Biggl\{\cos b_kx \int_{b_{k-}}^{b_{k+}}u^{-\mu}\cos u du+\nonumber \\
&&+\sin b_kx\Biggl[\int_0^{b_{k-}}u^{-\mu}\sin u du+ \int_0^{b_{k+}}u^{-\mu}\sin u du\Biggr]\Biggr\}, \nonumber \\
&&b_k=k\pi,\ b_{k\pm}=b_k(1\pm x). \label{so21a}
\end{eqnarray}
For $\mu=1$ we have from \eqref{so21a}
\begin{eqnarray}
&&g_{k 1}(x)=k\Biggl\{\sin b_k x\Bigl({\rm{Si}}\ b_{k-}+{\rm{Si}}\ b_{k+}\Bigr)-\nonumber \\
&&-\cos b_k x\Bigl({\rm{Ci}}\ b_{k-} - {\rm{Ci}}\ b_{k+}\Bigr) \Biggr\}.\label{so23}
\end{eqnarray}
We find analytically for $\mu=1$}}
\begin{equation}\label{so24}
\eta_{kk}=2k\ {\rm{Si}}(2k \pi).
\end{equation}
{{For $\mu=1$}} the solutions for the $2\times 2$ matrix have the form
\begin{eqnarray}
&&E_1=2.81019,\ E_3=5.99476,  \label{so25} \\
&&\psi (E_1)=(-0.995891,\ 0.0905574), \nonumber \\
&&\psi (E_3)=(0.0905574,\ 0.995891). \label{so26}
\end{eqnarray}
{{We note here that since the integrals $f_{k1}(x)$ and $g_{k1}(x)$ for $\mu=1$ can be expressed through known special functions Ci$(x)$ and Si$(x)$, which have very good polynomial approximations \cite{abr}, the calculations for this case are much faster (and much less computer intensive) then those for $\mu \neq 1$. That is why all calculations with very large matrices like 10000 $\times$ 10000 have been performed here for the case $\mu=1$, keeping in mind that the results for $\mu \neq 1$ behave themselves qualitatively similar with matrix size growth.}}

Two lowest eigenvalues of  the  $6 \times 6$ matrix {{for $\mu=1$}} read $E_1=2.78021,\ E_3=5.93979$. In Table  \ref{rt}  we reproduce the remaining four eigenvalues in the   $6\times 6$ case, in a comparative vein. Namely, we display the computation outcomes  for  lowest six  eigenvalues,  while  gradually increasing the matrix size, from  $6\times 6$, $12\times 12$ to  $10000\times 10000$. We reintroduce the traditional labeling in terms of $i=1,2,3,4,5,$   so that no explicit distinction is made between even and odd eigenfunctions. Our results are directly compared with the corresponding data obtained by other methods in Refs.  \cite{K,KKMS} and \cite{ZG,zg}.

In  Table \ref{t3} we report  the change of the  ground state energy while increasing the matrix size from $30\times 30$  to $10000\times 10000$. It is seen that the  third significant digit stabilizes already for $300\times 300$ and $400\times 400$ matrices.
\begin{figure}[tbh]
\centering
\includegraphics[width=0.9\columnwidth]{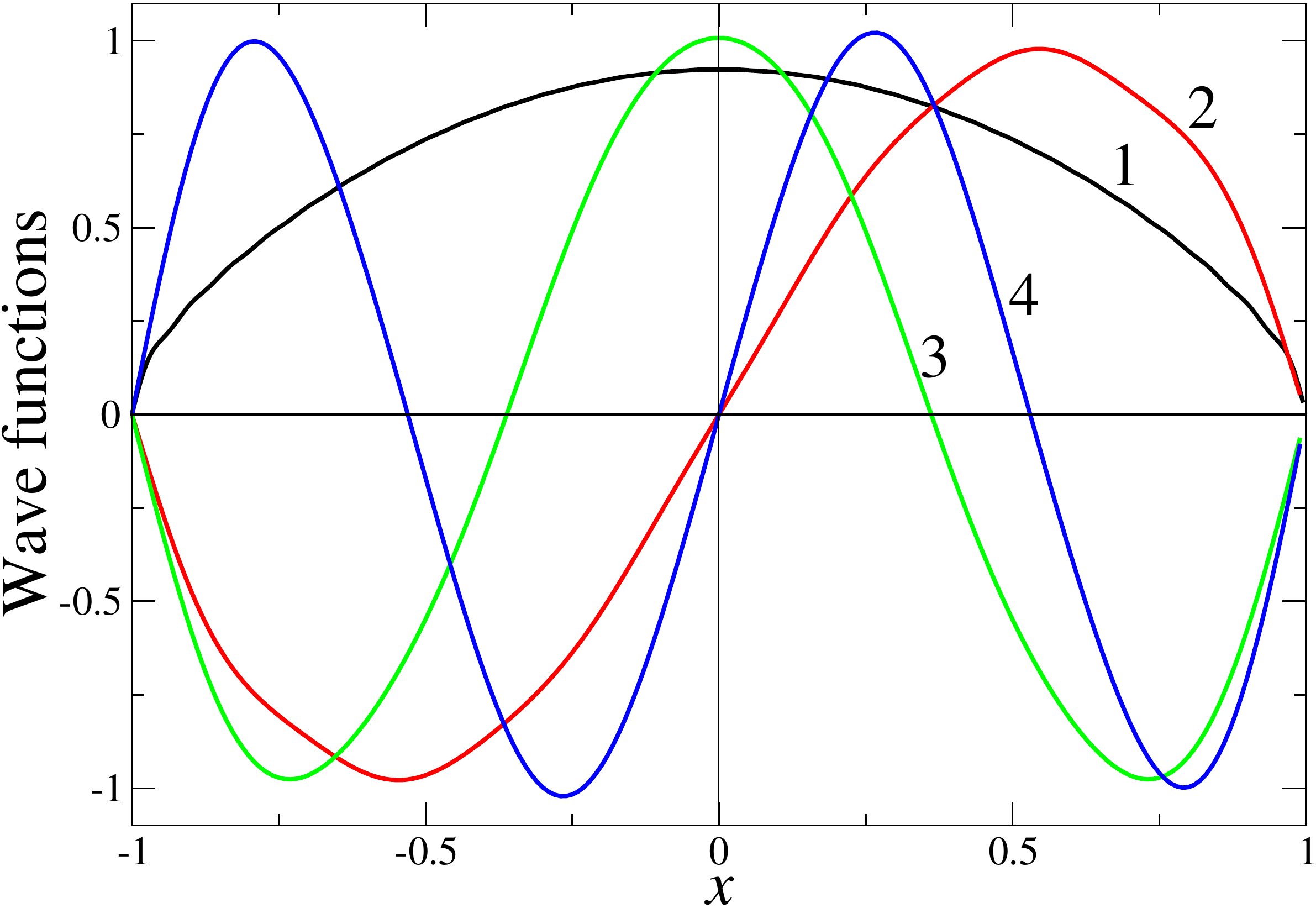}
\caption{(color online) Four lowest eigenfunctions in the infinite Cauchy well ($\mu=1$), labeled $i=1,2,3,4$. Outcome of  the  $10^4\times 10^4$ matrix. {{The qualitative behavior of the eigenfunctions for $\mu \neq 1$ is the same.}} } \label{fig3}
\end{figure}

\begin{figure}[tbh]
\centering
\includegraphics[width=1.1\columnwidth]{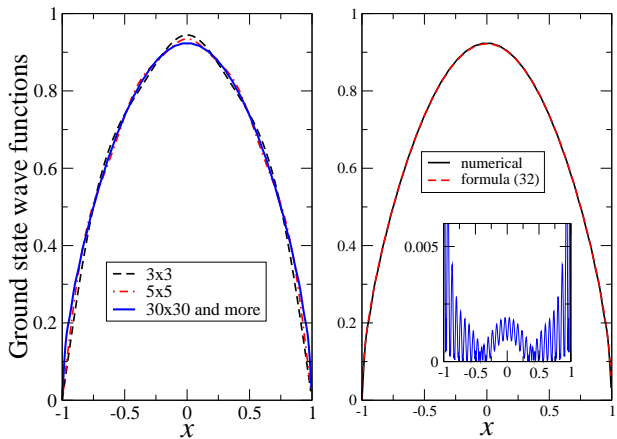}
\caption{(color online) Left panel. Comparison of the shapes of ground state  functions obtained by the diagonalization of 3 $\times$ 3 (black dashed curve), 5$\times$5 (red dash-dot curve) and 30$\times$30 (blue solid curve) matrices. The shape of ground state functions for matrices more then 30$\times$30 are identical to that for 30$\times$30. Right panel shows the approximation of ground state wave function (for 700$\times$700 matrix, solid curve) by the expression \eqref{zg} (dashed curve). As both lines are indistinguishable in the scale of the  figure, the inset depicts the modulus of the point-wise difference of respective curves} \label{fig4}
\end{figure}

\subsection{Graphical comparison}

We begin with plot of the first four eigenfunctions for {{representative value $\mu=1$, reported in Fig. \ref{fig3}. The situation for other $\mu$'s is qualitatively similar.}} It is seen that the states in the Cauchy well at a rough graphical resolution level resemble those of  the ordinary (deriving form the Laplacian) quantum  infinite  well \cite{landau3}. {{This speaks in favor of our statement that fractional Laplacian "mixes" the states, generated by ordinary one.}} The detailed analysis of the eigenfunctions shape issue can be found in Ref. \cite{zg}, where another method of solution of the Cauchy well  problem has been tested.

\begin{table*}
\begin{tabular}{|c|c|c|c|c|c|c|c|c|c|}
\hline
$n$ (matrix $n\times n$)  & 30 & 50 & 100 & 200 & 400 &1000 &
2000 & 5000 & 10000 \\ \hline
$E_{g.s.}=E_1$ &1.160505 & 1.159428 & 1.158608 & 1.158193 & 1.157984 & 1.157858
& 1.157816 & 1.157791 &1.157791 \\ \hline
$E_2$ &2.760953 & 2.758572 & 2.756705 & 2.755742 & 2.755252 & 2.754954
& 2.754855 & 2.754795 &2.754795 \\ \hline
$E_3$ & 4.326418 & 4.322736 & 4.319842 & 4.318343 & 4.317578 & 4.317114
& 4.316958 & 4.316864 &4.316864 \\ \hline
$E_4$ &5.904768 & 5.900041 & 5.896238 & 5.894235 & 5.893204 & 5.892573
& 5.892361 & 5.892233 &5.892233\\ \hline
$E_5$ & 7.476052 & 7.470114 & 7.465334 & 7.462812 & 7.461511 & 7.460714
& 7.460446 & 7.460284 &7.460284 \\ \hline
$E_6$ &9.051406 & 9.044604 & 9.039015 & 9.036021 & 9.034462 & 9.033504
& 9.033180 & 9.032984 &9.032984 \\ \hline
\end{tabular}
\caption{The matrix $n \times n$ - "size evolution" of six lowest eigenvalues {{for $\mu=1$}} as $n$ grows. $E_{g.s.}$ stands for ground state energy. }\label{t3}
\end{table*}

Since, in the present paper,  we employ trigonometric functions as  the orthonormal basis  system,
for low-sized matrices (\ref{so13})  we deal with visually distinguishable oscillations.
These are gradually smoothened with the growth of the matrix size. It is instructive to  compare approximate  shapes of the  ground state wave function, obtained by the diagonalization of  different-sized  matrices.
The left panel of Fig. \ref{fig4} reports the  pertinent  shapes in case of  $3\times 3$, $5\times 5$ and $30\times 30$ matrices for $\mu=1$. We note that the qualitative features of  the ground state function  approximants  are practically the same for matrices  of sizes exceeding $30\times 30$. {{This statement is also valid for general case $\mu \neq 1$.}}

In Ref. \cite{zg}, an analytic approximation of the ground state function  of $|\Delta |^{1/2}_D$ {{(i.e. that for $\mu=1$)}} has been proposed in the form
\begin{eqnarray}
&&\psi_1(x)=\psi_{g.s.}(x)=0.921749\sqrt{(1-x^2)\cos \alpha x},\nonumber \\
&&\alpha =\frac{1443\pi}{4096}. \label{zg}
\end{eqnarray}
In the right panel of Fig. \ref{fig4}, we  compare the ground state function \eqref{zg} with that obtained by the diagonalization of $700\times 700$ matrix  (which turns out to be close  to that obtained by  means of the  $30\times 30$ matrix, see Fig. \ref{fig5}).
It is seen that both functions are indistinguishable within the scale of the figure. The inset in Fig. \ref{fig4}  depicts the modulus of the point-wise difference of these functions. Interestingly, although the approximation is non monotonous (the difference oscillates),  in a  large portion of the interval
$-1\leq x\leq 1$ the difference does not exceed $0.005$.

\begin{figure}[tbh]
\centering
\includegraphics[width=1.1\columnwidth]{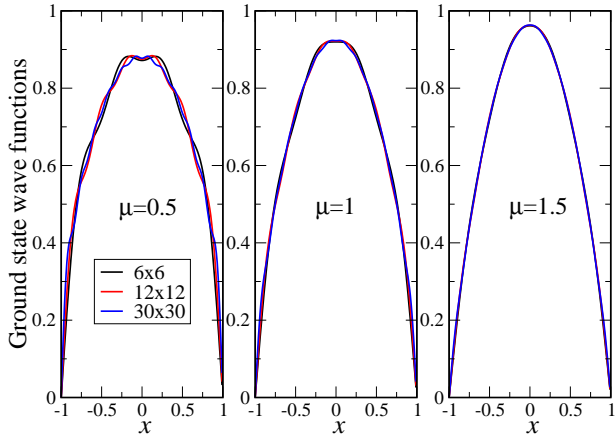}
\caption{(color online) {{Ground state wave functions for different L\'evy index $\mu$, obtained for $6\times 6$,
$12\times 12$ and $30\times 30$ matrices.}} } \label{zu1}
\end{figure}

{{The ground state wave functions for different $\mu$'s and obtained from the diagonalization of $6 \times 6$,
$12 \times 12$ and $30\times 30$ matrices are reported in Fig. \ref{zu1}. It seen that the closer $\mu$ to 2 (ordinary Laplacian), the faster is convergence. Namely, while for $\mu=1.5$ the outcome of the matrix $6 \times 6$ is to second decimal place is similar to that for $30\times 30$ matrix, in the case $\mu=0.5$ the difference is distinguishable in the scale of the figure. This fact shows that as $\mu \to 2$, the number of base functions, "taking part" in the wave function approximation (i.e. the order of the corresponding matrix) tends to only one, corresponding to that for ordinary quantum mechanical infinite well. }}

Generally, for the approximate eigenfunction,  the  function $ |\Delta |^{\mu/2}_D\psi(x)$ differs from $E\psi(x)$ and symptoms of  convergence are  expected with the growth of the matrix size. In  Fig. \ref{fig5} we compare the left- and right-hand sides of the integral equation \eqref{ii2a} {{for $\mu=1$ and show the}} modulus of their difference.

\begin{figure}[tbh]
\centering
\includegraphics[width=0.95\columnwidth]{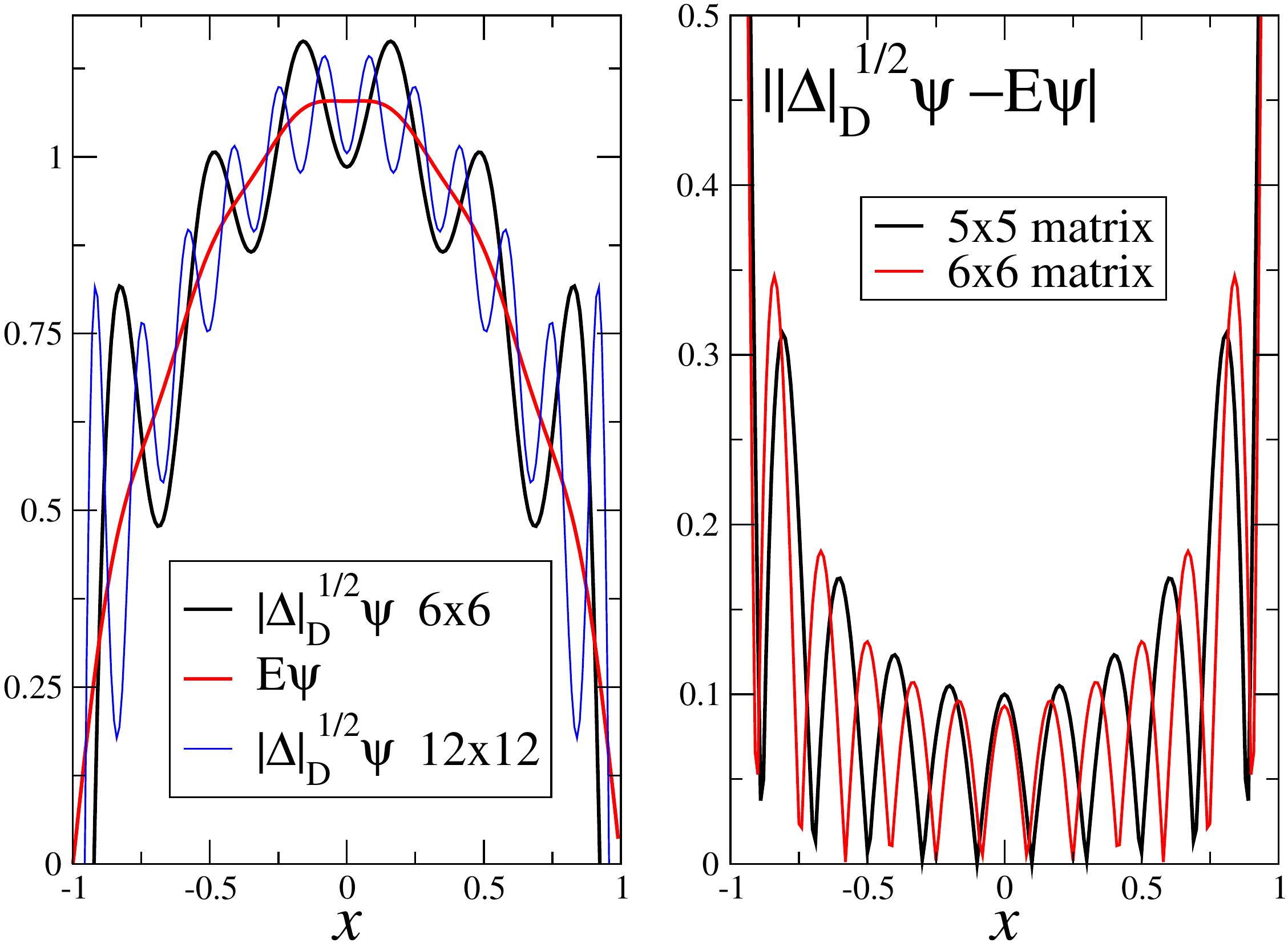}
\caption{(color online) Left panel. Comparison of $|\Delta |^{1/2}_D\psi$ (black curve) and $E \, \psi$ (red curve) for $6\times 6$ matrix. Thin blue line corresponds to $12\times 12$ matrix. It is seen that for the $12\times 12$ matrix $ |\Delta |^{1/2}_D\psi$ goes closer to $E\psi$ in the main body of the interval.
Right panel. The deviation $||\Delta |^{1/2}_D\psi - E\, \psi | $ for $5\times 5$ (black curve) and $6\times 6$ matrices (red curve).} \label{fig5}
\end{figure}

\begin{figure}[tbh]
\centering
\includegraphics[width=0.95\columnwidth]{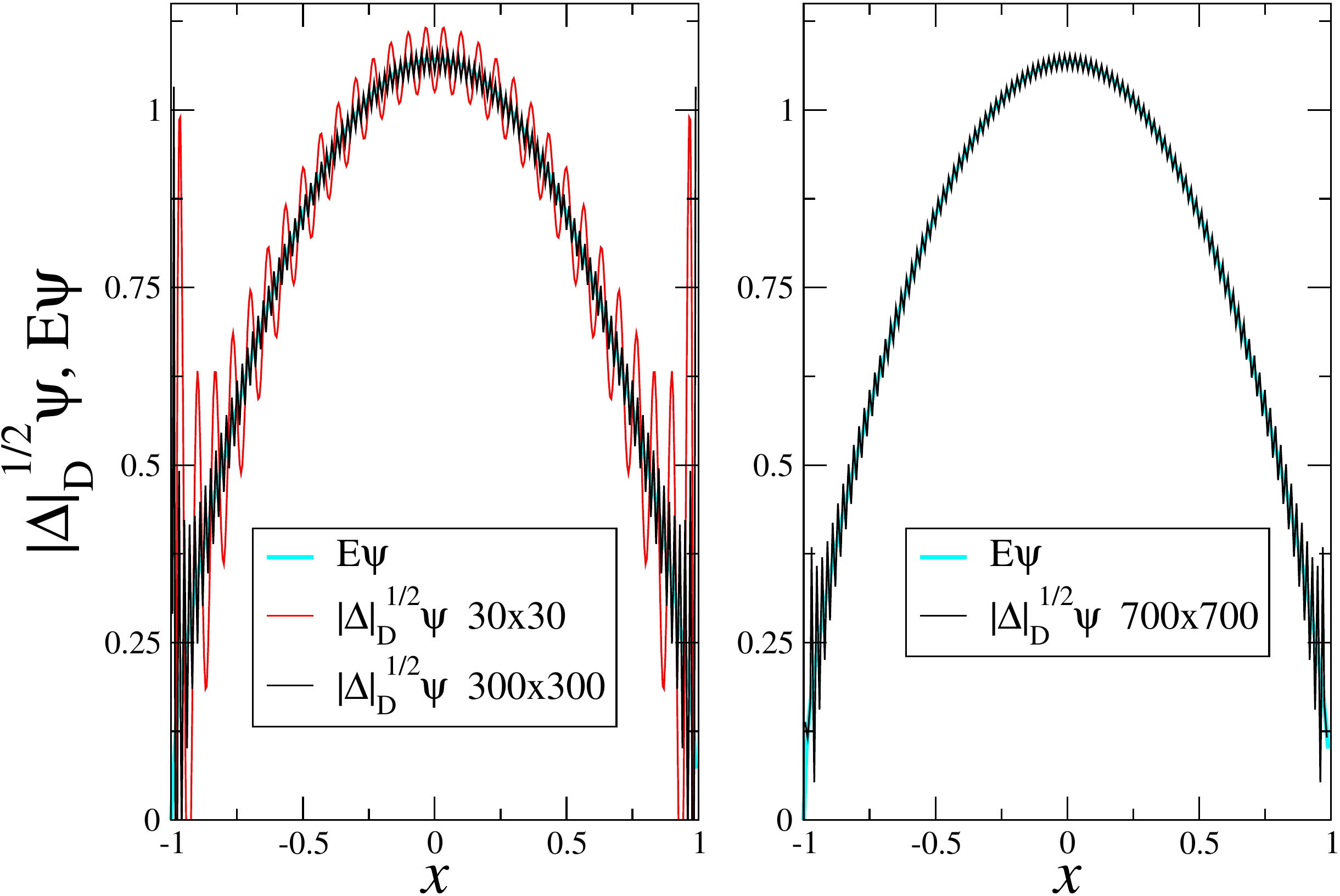}
\caption{(color online) More detailed comparison of $E\psi$ (cyan curve) and $|\Delta |^{1/2}_D\psi$ (red and black curves) for $30\times 30$
 (red curve) and $300\times 300$ (black curve) matrices (left panel). Right panel shows that for matrices larger then $300\times 300$,  the "wavy approximation" of $\psi(x)$ approaches (in the adopted scale) the line thickness everywhere,  except the  close vicinity of $x=\pm 1$ points.} \label{fig6}
\end{figure}

Left panel of Fig. \ref{fig5} shows that while the  function $\psi$ by itself is smooth,
the function $|\Delta |^{1/2}_D\psi$ is "wavy" and diverges at the boundaries.
 The right panel shows the error $||\Delta |^{1/2}_D\psi-E\psi|$ for $6\times 6$ and $12\times 12$ matrices.
  It is seen that the divergence at the boundaries is  qualitatively  the same for both cases,
  while for the $12\times 12$ matrix the error in the vicinity of  $x=0$ is a little smaller. {{The same tendence occurs at any $0<\mu \leq 2$.}} This kind of behavior (slow convergence {{of $|\Delta |^{\mu/2}_D\psi$ to $E\psi$)}} is characteristic for integral equations with singular kernels \cite{pol} and also for "ordinary" quantum mechanical  spectral  problems,  if  we solve them approximately by  the  expansion method with respect to  the full set of eigenfunctions of another operator (here, the Laplacian $\Delta $).

Figure \ref{fig6} reports the behavior of $E\psi$ and $ |\Delta |^{1/2}_D\psi$ for $30\times 30$ and $300\times 300$ matrices. The "wavy"  behavior of  $A_D\psi$ persists, while  $E\psi$ stabilizes  already beginning from the  12x12 matrix. Our analysis shows that if we take larger matrices, the diverging   "tail"  moves  closer to boundary points  $\pm 1$ so that at $n \to \infty$ ($n$ is order of the matrix) it disappears.
 The same is valid for superimposed  oscillations, whose  amplitude (slowly) diminishes as $n$ grows. Similar to the situation in Fig. \ref{zu1}, the good convergence for smaller $\mu$'s is achieved for larger $n$.
 On the other hand, even for relatively small matrices $6\times 6$ we obtain qualitatively reasonable
approximations for eigenfunctions and eigenvalues of the {{operator \eqref{cw2}, especially for indices $\mu$ close to 2.}}

\begin{table}
\begin{tabular}{|c|c|c|c|c|}
\hline
$n$ & $E_{n, 5000 \times 5000}$ & Eq.\eqref{kw} & Rel. error, \% & Data from \cite{KKMS} \\ \hline
1 &1.157791 & 1.178097 & 1.75 & 1.157773 \\ \hline
2 & 2.754795 & 2.748894 & 0.21 & 2.754754 \\ \hline
3 & 4.316864 & 4.319690 & 0.06 & 4.316801   \\ \hline
4 & 5.892233 & 5.890486 & 0.03 &  5.892147 \\ \hline
5 & 7.460284 & 7.461283 & 0.013 &  7.460175  \\ \hline
6 & 9.032984 &  9.032079 & 0.01 &   9.032852 \\ \hline
7 & 10.602447 & 10.602875 & 0.004 & 10.602293   \\ \hline
8 & 12.174295 & 12.173672 & 0.0051 & 12.174118 \\ \hline
9 & 13.744308 & 13.744468 & 0.0012 &  13.744109 \\ \hline
10 & 15.315777 & 15.315264 & 0.0033 &  15.315554 \\ \hline
11 & 16.886062 & 16.886061 & 5.9$\cdot 10^{-8}$ & * \\ \hline
12 & 18.457329 & 18.456857 & 0.0026 & * \\ \hline
13 & 20.027767 & 20.027653 & 0.00057& * \\ \hline
14 & 21.598914 & 21.598449 & 0.0021 & *\\ \hline
15 & 23.169448 & 23.169246 & 0.00087 & * \\ \hline
16 & 24.740517 & 24.740042 & 0.0019  &  * \\ \hline
17 & 26.311115 & 26.310838 & 0.0011 &  * \\ \hline
18 & 27.882131 & 27.881635 & 0.0018 &  * \\ \hline
19 & 29.452773 & 29.452431 & 0.0012 &  * \\ \hline
20 & 31.023751 & 31.023227 & 0.0016 &  * \\ \hline
30 & 46.731898 & 46.731191 & 0.0015 &  * \\ \hline
50 & 78.148251 & 78.147117 & 0.0015 & * \\ \hline
100 & 156.689159 & 156.686934 & 0.0014 & * \\ \hline
\end{tabular}
\caption{The comparison of several eigenvalues of the $5000 \times 5000$ matrix (\ref{so13}) {{for $\mu=1$}} with the  approximate formula  $n\pi /2 - \pi /8$ {{(Eq. \eqref{kw} at $\mu=1$)}} along with the relative error $|E_n- (n\pi /2 - \pi /8)|/E_n$. Independently obtained spectral  data (formula (1.11) in \cite{KKMS}) are displayed as well.}\label{kuka}
\end{table}

If compared with the previous methods of solution \cite{KKMS,K} and \cite{ZG,zg}, our spectral approach seems to be particularly powerful if one is interested in the spectrum {{of $|\Delta |^{\mu/2}_D$.}} In fact, we are able to generate an  arbitrary number of eigenvalues and corresponding eigenfunctions with any desired accuracy.
 In Table \ref{kuka} we compare several (first 20 and  a couple of larger) lowest eigenvalues {{of $|\Delta |^{1/2}_D$ (i.e. for typical case $\mu=1$)}} and answer how much actually  {{the approximate formula  \eqref{kw}  deviates}} from  computed  $E_n$'s.

It is seen from the Table \ref{kuka} that although the asymptotic formula delivers pretty good approximation to the desirable eigenvalues, the relative error never (except for $n=11$)  falls below $10^{-3}$ \%
as the label  number $n$ grows.  We  have actually   traced  this statement  up to $n= 500$.
Moreover, the relative error, as it is seen from the Table \ref{kuka}, oscillates around $10^{-3}$ \%, which means that beginning with $n \approx 8$ the expression \eqref{kw} for $\mu=1$ contributes 5 significant digits of the "true" asymptotic  answer. {{Note that for $1< \mu \leq 2$ this number $n$ diminishes so that at $\mu=1.9$ the same result is obtained already for $n=2$. On the other hand, at $\mu=0.2$ for $n=50$  we have only 3 significant digits.}}

\begin{figure*}
\includegraphics[width=0.95\columnwidth]{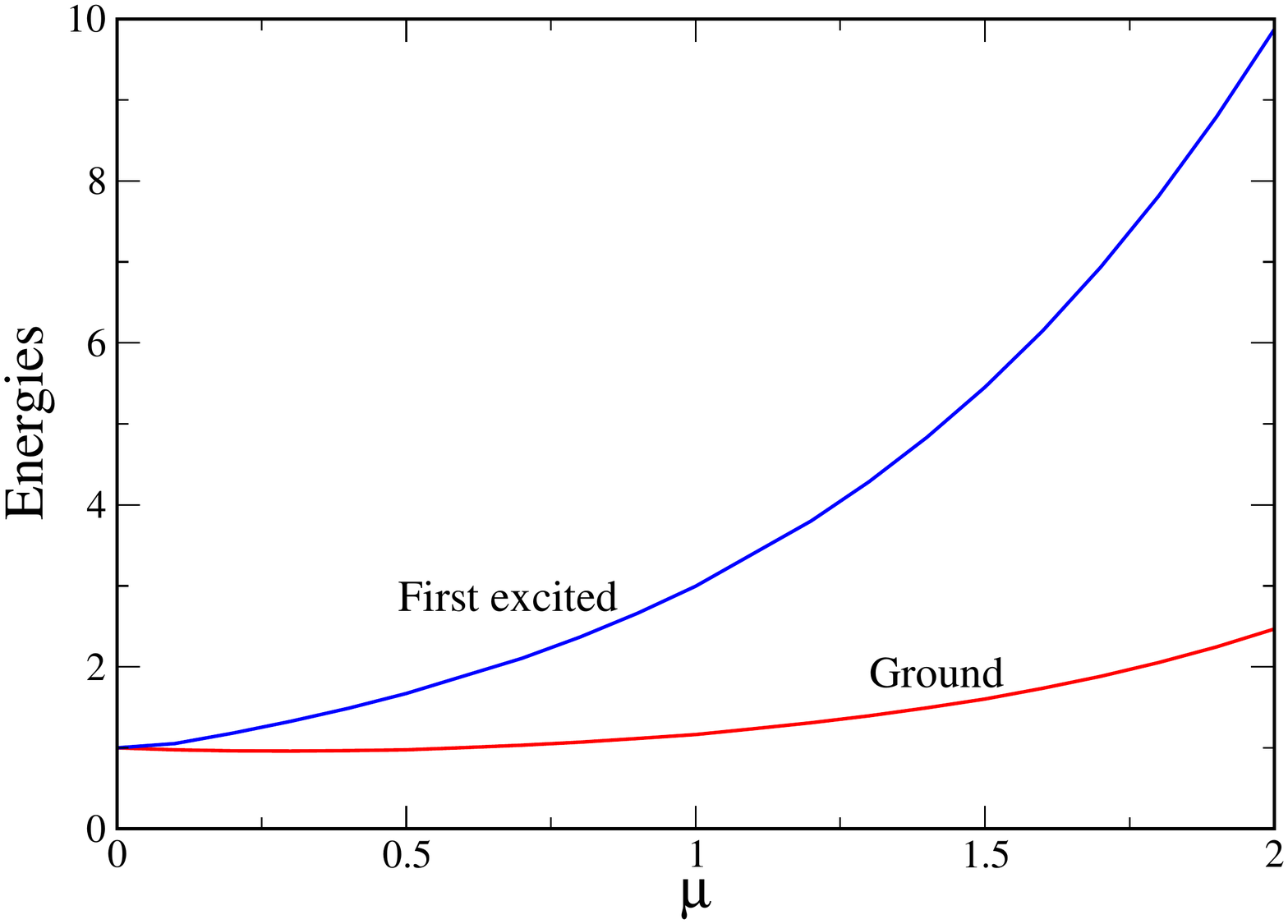}
\hspace*{-5mm}
\includegraphics[width=0.95\columnwidth]{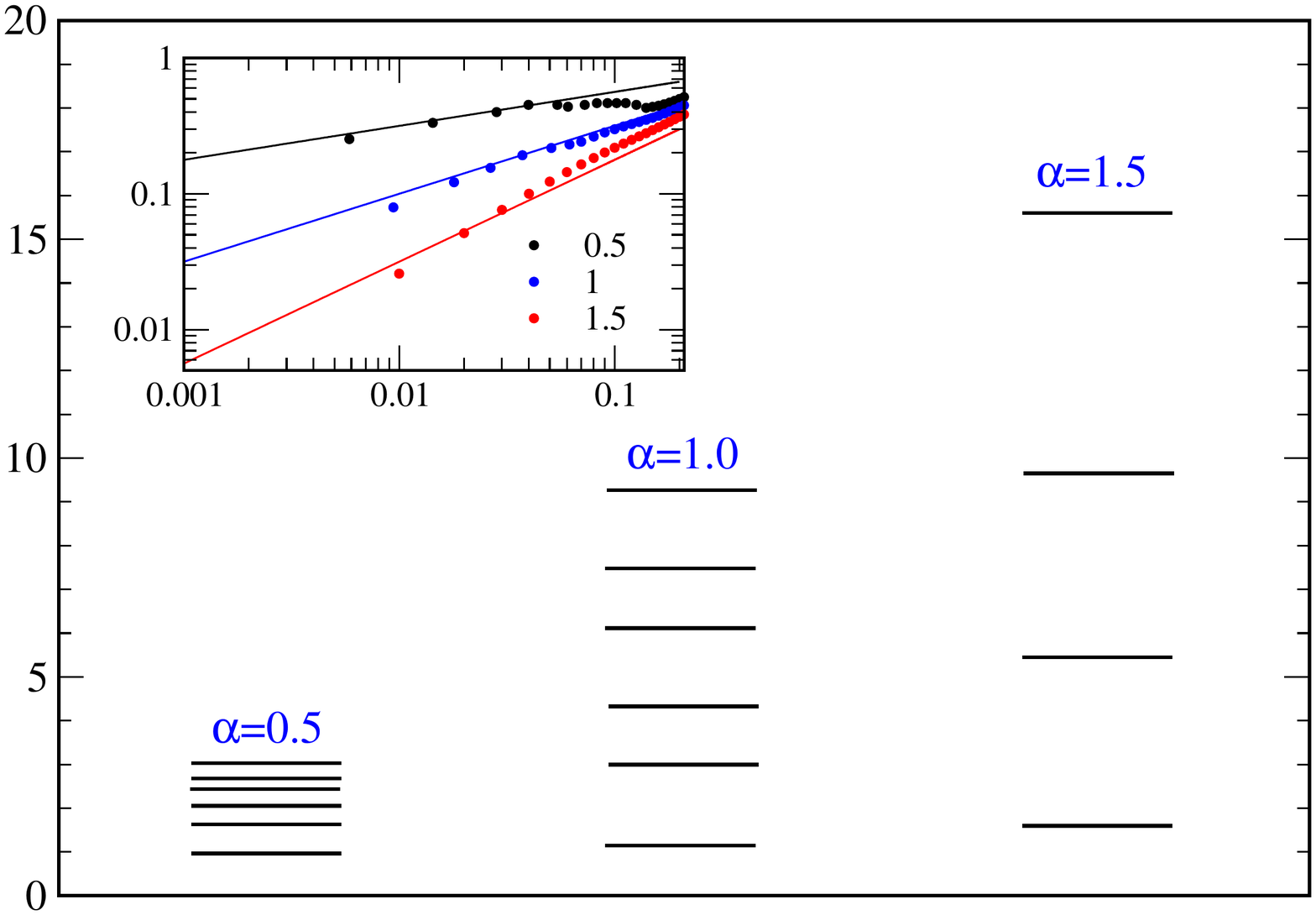}
\caption{(color online) The properties of the spectrum of integral equation \eqref{ii2a} for different $\mu$'s.
Left panel shows the dependence of ground and first excited states energies on the parameter $\mu$.
At $\mu=2$ the energy levels positions are equal to those in the ordinary quantum well \eqref{ord}. At $\mu=0$ all spectrum merges into the single level $E_0=1$ (in our units). Right panel visualizes how several first energy levels look like for different $\mu$'s. The shrinking of the spectrum at $\mu \to 0$ is clearly seen. The inset shows in double log plot the character of decay of ground state wave functions at $x=\pm 1$. For better visualization we shift the left part of the functions (i.e. those at $x=-1$) to zero. Points report the wave functions and lines are $(x+1)^{\mu/2}$ for $\mu=$ 0.5, 1 and 1.5 respectively.} \label{qu}
\end{figure*}

{{Although the numerical calculations for the case $\mu=1$ are (sometimes much) less computer intensive then those for $\mu \neq 1$, the former case does not permit to trace the additional properties of the spectrum of the equation \eqref{ii2a}, which depend on $\mu$. These properties are summarized in Fig. \ref{qu}.
In left panel we plot the $\mu$ dependence of ground and first excited state energies. All other energies are also avalilable but their values at $\mu=2$ grow rapidly with $n$ so that for higher excited states not all $\mu$'s will fit  the scale of the plot. This is because at $\mu=2$ the spectrum of the equation \eqref{ii2a} gives exactly that for ordinary quantum well \cite{landau3}, which in our units has the form
\begin{equation}\label{ord}
E_{n,\mu=2}=\frac{\pi^2}{4}n^2,\ n=1,2,3,...,
\end{equation}
i.e. it is proportional to $n^2$. Note that at $\mu=2$ the approximate dependence \eqref{kw} yields exactly \eqref{ord} thus also giving exact known result.

Substituting $n=1$ and 2 into Eq. \eqref{ord} we have, respectively $E_{1,\mu=2}=\pi^2/4\approx 2.4674$ and
$E_{2,\mu=2}=\pi^2\approx 9.8696$, which are seen in the left panel of Fig. \ref{qu} at $\mu=2$. For instance, at $n=3$ $E_{3,\mu=2}=9\pi^2/4 \approx 22.2066$, which is two times larger then $E_{2,\mu=2}$.

The most interesting feature of our method is that it permits to transit smoothly to the case $\mu=0$, which does not included in the domain of the operator \eqref{cw2}. Moreover, the integrals \eqref{so4b} and \eqref{so21a} can be exactly evaluated in this case. This gives explicitly
\begin{equation}\label{mu0}
f_{k,\mu=0}=\cos \lambda_kx,\  g_{k,\mu=0}=\sin b_kx,
\end{equation}
which, in turn, yields
\begin{equation}\label{mu1}
\gamma_{\mu=0, ki}=\eta_{\mu=0, ki}=\delta_{ki}.
\end{equation}
The expression \eqref{mu1} immediately shows that at $\mu=0$ all eigenvalues of Eq. \eqref{ii2a} equal to 1. In other words, the entire spectrum of the operator \eqref{cw2} at $\mu=0$ shrinks into one single value $E_0=1$.
This value is seen on the left panel of Fig. \ref{qu}.

The character of spectrum shrinking at $\mu \to 0$ is shown on the right panel of Fig. \ref{qu}. Here we report several first energy levels for different $\mu$'s. The expansion of the spectrum as $\mu \to 2$ and its shrinking as $\mu \to 0$ is clearly seen.

An important feature of the eigenfunctions of the operator \eqref{cw2} is that, contrary to trigonometric functions for $\mu=2$, they decay nonlinearly at $x=\pm 1$. The hypothesis is that they vanish as $(1\pm x)^{\mu/2}$. To check this hypothesis, in the inset to lower panel of Fig. \ref{qu} we plot (shifting for convenience the left edge $x=-1$ to zero) the ground state wave functions (points) for different $\mu$'s along with functions $(1+x)^{\mu/2}$ in double logarithmic scale (full lines). It is seen that at $0.01 < x < 0.1$ the coincidence is almost perfect. However, small deviations are seen already at $x=0.01$. This is related to the approximate character of wave functions. To continue the points to smaller $x$, the consideration of larger matrices is necessary, which (even for eigenvectors of large matrices, corresponding to $\mu=1$) is extremely computer intensive task. Most probably, the hypothesis about asymptotics $(1\pm x)^{\mu/2}$ is true.}}

\section{Conclusions}

In the present paper we have studied the spectrum of the problem of a particle in the infinite potential well, obeying fractional quantum mechanics with arbitarary L\'evy index $0<\mu \leq 2$. This problem is relevant to many disordered and dissipative physical (and biological, chemical and even social) systems, involving L\'evy flights in bounded domains and is nontrivial as the familiar representation of fractional derivatives in Fourier domain does not work in such confined case. To solve this problem, we reduce the initial fractional
Scr\"odinger equation to the Fredholm integral equation with hypersingular kernel. For the solution of latter equation, we have elaborated a novel and powerful method, based on the expansion over the complete set of the {{orthogonal functions taken from corresponding "ordinary" quantum mechanical problem for L\'evy index $\mu=2$.}} In our case of the interval $-1\leq x \leq 1$ we use the trigonometric functions, which are eigenfunctions of the ordinary Laplacian. {{We note here the general character of our method in the sense of its applicability to virtually any "ordinary" quantum mechanical problem, also in two and three dimensions.

Let us finally mention the realistic physical systems, where we are going to apply our formalism. One of the important example is electronic tunneling characteristics in spintronic devices \cite{zu,ya,ng}. Spintronics or spin electronics is nowadays a branch of physics whose central theme is the active manipulation of spin degrees of freedom in solid-state systems \cite{zu,cn,ds}. It is widely believed, that spintronic devices can lead to applications (like quantum computers) that are so far infeasible with modern electronics. Despite intense experimental and theoretical studies, the statistics of tunneling electrons through barriers in such structures remains unclear due to disorder which is inevitably present in such structures \cite{cn,ds}. The above barriers are technologically realized in inversion semiconductor layers, heterostructures (like perovskite interface LaAlO$_3$ - SrTiO$_3$ \cite{oh,sd1,sd2}), quantum wells or in graphene \cite{cn,ds,gn}. The common formalism for description of electronic states in the above structures is different variations of particle in a potential well problem (see, e.g. Ref. \cite{sze} and references therein). It has been suggested that to describe the tunneling statistics adequately, the fractional derivatives should be brought into the above formalism. We are going to apply the developed formalism to the "disordered" quantum wells as well as to oxide interfaces \cite{oh,sd1}, which also has interesting and non-trivial physical properties. One more problem is the electronic properties of so-called multiferroics, i.e. substances, combining several types of long-range orders (like ferroelectricity and ferromagnetism). The non-Gaussian statistics due to disorder plays an important role in these substances also \cite{lag1,lag2} and we are applying now our formalism (also in context of the problem of quantum oscillator with fractional Laplacian) to obtain the adequate description of their physical properties.}}

\end{document}